\documentstyle[12pt]{article}
\newcommand{\be}{\begin{equation}}
\newcommand{\ee}{\end{equation}}
\newcommand{\bea}{\begin{eqnarray}}
\newcommand{\eea}{\end{eqnarray}}
\newcommand{\bean}{\begin{eqnarray*}}
\newcommand{\eean}{\end{eqnarray*}}
\newcommand{\ba}{\begin{array}{l}}
\newcommand{\ea}{\end{array}}
\newcommand{\bb}{\begin{thebibliography}{99}}
\newcommand{\eb}{\end{thebibliography}}

\newcommand{\Tr}{\mbox{Tr\,}}
\newcommand{\Ds}{\displaystyle}

\begin{document}

$\left.\right.$
\vskip 2cm
\Large
\begin{center}
{\bf On the effective action\\
of local composite fields }
\end{center}
\large
\centerline{ S.A.Garnov }
\centerline{ Department of Physics, Institute of }
\centerline{ Transport Engineering, Tashkent, Uzbekistan }
\vskip 1cm

\normalsize
The gauge invariant method for calculation of the effective action
of the local composite fields in QFT is proposed. The effective action
of the local composite fields in QED is studied up to 2-loop level.
The graph rules for the local composite fields are derived. On the
basis of these rules the problem of one-particle irreducibility is
discussed.

\vskip 1cm

\large
1. Composite fields have their applications in various domains of the
theoretical physics (see, for example, [1] and references there). In order
to study the composite fields in the framework of the functional approach
we should calculate the corresponding generating functionals. The effective
action is one of such functionals which is particularly usefull to derive
the dynamical information from a system.

The effective action was thoroughly studied for the  case of non-local
composite fields in [2], [3]. Another interesting problem is the
investigation of the gauge-invariant local composite fields. There were
attempts to use well-studied effective action of non-local composite fields
as a starting point when study the local composite fields ([4], [5])
but this is not quite correct because the discussed approaches were not
gauge-invariant by themselves [6]. It is important, therefore, to develop
a special method for calculation of the effective action of local composite
fields which preserves the gauge-invariance.

The existing methods of calculating the effective action ([6], [7], [8])
applied to composite fields lead to substantial difficulties because of
the necessity to sum over various classes of graphs corresponging to the
chosen composite field. The "classic" composite field which is the functional
argument in the effective action can be derived this way. This program is
implemented in [6] in the framework of "inversion" technic.

It is obvious, that in order to avoid the summing over graphs we should work
with such equations that contain the full expressions of "classic" field, not
its series representation. Becides, for the reason of investigation of the
one-particle irreducibility (1PI) we should completely remove the explicit
dependence on sources from the effective action.

2. We want to propose a method of calculation of the effective action which is
complied with the above-mentioned conditions. As an example we will calculate
the effective action of a local composite field, namely EM current in QED.

The first step will be the calculation of the effective action for the ordinary
field. The obtained expression will be used to receive the equation for the
composite field. Having resolved the last one we will get the effective action
for the local composite field.

We shall follow here the work [9] and only at the end
of this section we make some generalisations on the calculating traces
of $\gamma$- matrixes which are important
for higher-loop calculations.

The generating functional for the spinor electrodynamics with Lagrangian

\be L = -\frac{1}{4} F_{\mu \nu} F^{\mu \nu} + \bar{\psi}
[ i(\gamma^{\mu} \partial_{\mu} - ie \gamma^{\mu} A_{\mu}) + m]\psi -
\frac{1}{2\alpha}(\partial_{\mu}A^{\mu})^{2} \ee
has the form (the source is introduced only for the photon field)
\be Z[J_{\mu}] = \int DA D\psi D\bar{\psi} exp [i\int L dx + i(JA)] \ee
Having integrated over spinor variables we get
\be Z[J_{\mu}] = \int DA  \exp [i S_{eff} + i(JA)] \ee
where
\be S_{eff} = S_{A} + i \hbar T \ee
\be \ba \Ds S_{A} = \frac{1}{2} A_{\mu} D^{-1}_{\mu \nu} A_{\nu} , \;
T = \Tr \ln K , \\
\Ds K^{-1} = i\gamma^{\mu} \partial_{\mu} + e \gamma^{\mu} A_{\mu} - m,
\; D_{\mu \nu} = \frac{1}{\partial^{2}} (g_{\mu \nu} - (1 - \alpha)
\frac{\partial_{\mu} \partial_{\nu}}{\partial^{2}}) \ea \ee

Let us define the effective action in the following way
\be \Gamma[\langle  A_{\mu}\rangle] = W[J_{\mu}] - (J_{\nu}
\langle  A^{\nu}\rangle) \ee
where
\be \langle A_{\mu}\rangle = Z^{-1} \int DA A_{\mu} \exp [i S_{eff} + i(JA)]
\ee
is so-called "classical field" and
\be W[J_{\mu}] = -i \ln Z \ee
The effective action defined this way is the generating functional of
the Green's functions which are 1PI with respect to photon propagator.
>From (3) and (6) we have
\be \frac{\delta \Gamma}{\delta \langle A_{\mu}\rangle} =
\langle \frac{\delta S_{eff}}{\delta A_{\mu}}\rangle \ee

Further we will use the DeWitt's approch [8].
For the models containing fermion and vector fields the general form of
DeWitt's formula is
\be \langle Q[A,\bar{\psi},\psi]\rangle =
:\exp(\hat{G}): Q[\langle A\rangle,\langle \bar{\psi}\rangle,\langle
\psi\rangle]
\ee
where
\be \ba \Ds \hat{G} = \frac{i}{\hbar} \sum_{n=2}^{\infty}
\frac{(-i\hbar)^{n}}{n!}
\sum C_{ijk}^{n} (-1)^{j}
G^{\mu_{1}\ldots \mu_{i} \alpha_{1}\ldots\alpha_{j}\beta_{1}\ldots\beta_{k}}
\times \\ \Ds \times
\frac{\delta^{n}}{\delta\langle A_{\mu_{1}}\rangle\ldots \delta\langle
\psi_{\beta_{k}}\rangle
\ldots \delta \langle \bar{\psi}_{\alpha_{j}}\rangle\ldots} \ea \ee
\be
G^{\mu_{1}\ldots \mu_{i} \alpha_{1}\ldots\alpha_{j}\beta_{1}\ldots\beta_{k}}
= \frac{\delta^{n} W}
{\delta J_{\mu_{1}} \ldots \delta \eta_{\beta_{k}}
\ldots \delta \bar{\eta}_{\alpha_{j}} \ldots}  \ee
are the connected Green's functions with i photon and j+k fermion legs;
the colons mean that derivatives act only on $Q$; the Plank constant $\hbar$
is restored and
\be C_{ijk}^{n} = \frac{n!}{i! j! k!}, \; i+j+k=n \ee
Having applied (10) to the rhs of (9) we get
\be \frac{\delta \Gamma}{\delta \langle A_{\mu}\rangle} =  :\exp(\hat{G}):
\frac{\delta S_{eff}[\langle A_{\mu}\rangle]}{\delta \langle A_{\mu}\rangle}
\ee
In our case we have for $\hat{G}$
\be \ba \Ds \hat{G}=
\frac{-i\hbar}{2}G^{\mu\nu}\frac{\delta^2}{\delta \langle A^{\mu}\rangle
\delta \langle A^{\nu}\rangle} -
\frac{\hbar^{2}}{6}G^{\mu\nu\lambda}\frac{\delta^3}
{\delta \langle A^{\mu}\rangle \delta \langle A^{\nu}\rangle \delta \langle
A^{\lambda}\rangle} + \ldots,
\ea \ee
where
\be
G^{\alpha_{1} \ldots  \alpha_{n}}
= \frac{\delta^{n} W}
{\delta J_{\alpha_{1}} \ldots \delta J_{\alpha_{n}}} \ee

The effective action for the field $\langle A \rangle$
can be derived from this equation. Up to the 2-loop level the effective action
was obtained in [9]:
\be \Gamma_{ord} = \frac{1}{2} \langle A_{\mu}\rangle D^{-1\;\mu\nu} \langle
A_{\nu} \rangle
+ i\hbar T + \frac{1}{2}\hbar^{2} \Tr ( - D^{\mu\nu} T_{\mu\nu} ), \ee
where
\be
T^{\alpha_{1} \ldots  \alpha_{s}}
= (-1)^{s} \frac{\delta^{s} T}
{\langle \delta A_{\alpha_{1}}\rangle \ldots \delta \langle A_{\alpha_{s}}
\rangle} \ee

The condensed notations used up to now
imply summarizing over discrete variables and integrating over continuos
ones in all the expressions. Nevertheless the question of how to calculate
traces over $\gamma$-matrices was out of discussion. Therefore the method
described above should be expanded with the appropriate rule in order to
satisfy the common rules of diagrammatic technique. Namely, the calculations
of traces must be carried out along every fermionic loop separately ( in our
case the loops are constructed not out of the simple propagators but out
of the propagators in "external field" ).

3. Now we start investigating the composite fields. We define
the generating functional as following
\be \Gamma[\Phi] = W[J] - \int dx J_{\mu}(x)\Phi^{\mu}(x), \ee
where
\be \ba \Ds W[J] = -i\hbar \ln \int DA D\bar{\psi} D\psi
\exp [\frac{i}{\hbar} \int L dx
+ \frac{i}{\hbar}J_{\mu}(\bar{\psi} \gamma^{\mu} \psi)], \ea \ee
\be \Phi^{\mu}(x) = \frac{\delta W [J]}{\delta J} = \langle \bar{\psi}
\gamma^{\mu} \psi \rangle_{J}. \ee

Let us introduce an auxiliary functional
\be \ba \Ds \tilde{W}[J, K] = -i\hbar \ln \int DA D\bar{\psi} D\psi
\exp [\frac{i}{\hbar} \int L dx
+ \frac{i}{\hbar}J_{\mu}(\bar{\psi} \gamma^{\mu} \psi) + \frac{i}
{\hbar}(K_{\nu} A^{\nu})], \ea \ee

Having integrated over spinor variables we have
\be \ba \Ds \tilde{W}[J, K] = -i\hbar \ln \int DA \exp [\frac{i}{\hbar} S_{J}
+ \frac{i}{\hbar}(K_{\nu} A^{\nu})], \ea \ee
where
\be S_{J} = \frac{1}{2} A_{\mu} D^{-1}_{\mu \nu} A_{\nu} + i \hbar \Tr \ln
(i\gamma^{\mu} \partial_{\mu} + e \gamma^{\mu} A_{\mu} - m + \gamma^{\mu}
J_{\mu}) \ee
and $D_{\mu \nu}$ was defined in (5).
It is obvious that
\be \tilde{W}[J, K=0] = W[J]. \ee

Let us carry out the Legendre transformation on $K$ considering the
source of composite fields $J$ as a parameter:
\be \Ds \tilde{\Gamma}[\langle A \rangle _{K;J}; J] = \tilde{W}[K;J] - \int
dx K_{\mu}
\langle A^{\mu} \rangle _{K;J} \ee
\be \langle A^{\mu} \rangle _{K;J} = \frac{\delta \tilde{W}[J;K]}{\delta
K_{\mu}}. \ee

Further we will use the result of (17) keeping in mind the dependence on $J$:
\be \ba \Ds \tilde{\Gamma}[\langle A \rangle _{K;J}; J] =  \frac{1}{2} \langle
A_{\mu}\rangle _{K;J}
D^{-1\;\mu\nu} \langle A_{\nu} \rangle _{K;J} - \\
\Ds - i \hbar \Tr \ln [i\gamma^{\mu} \partial_{\mu} +
e \gamma^{\mu} \langle A_{\mu} \rangle_{K;J} - m + \gamma^{\mu} J_{\mu}] + \\
\Ds + \frac{\hbar^{2}}{2} \Tr \{ D^{\mu\nu} \frac{\delta^{2}}{\delta \langle
A^{\mu} \rangle
\delta \langle A^{\nu} \rangle} (\Tr \ln [i\gamma^{\mu} \partial_{\mu} +
e \gamma^{\mu} \langle A_{\mu} \rangle_{K;J} - m + \gamma^{\mu} J_{\mu}])\}.
 \ea \ee

Now let us put the sources of photon fields equal to zero. From (25)
and (26) we get
\be \Ds \tilde{\Gamma}[\langle A \rangle _{K=0;J}; J] = W[J], \ee
where $W[J]$ has been defined in (20). We can see that the equation (28)
expresses $W[J]$
as functional of $J$  and
$\langle A \rangle _{J} = \langle A \rangle _{K=0;J}$.

Further we should represent $\langle A \rangle _{J}$ in terms of $\Phi$.
With the integrations
over spinor variables restored we have
\be \ba \Ds \langle A_{\mu} \rangle _{J} = \frac{ \int DA D\bar{\psi}
D\psi A_{\mu}\exp [\frac{i}{\hbar} \int L dx
+ \frac{i}{\hbar}J_{\mu}(\bar{\psi} \gamma^{\mu} \psi)]}
{ \int DA D\bar{\psi} D\psi \exp [\frac{i}{\hbar} \int L dx
+ \frac{i}{\hbar}J_{\mu}(\bar{\psi} \gamma^{\mu} \psi)]}. \ea \ee

As the measure in the continual integral is invariant in respect to
small translations [10]
we have
\be \Ds \int DA D\bar{\psi} D\psi \frac{\delta}{\delta A^{\mu}}\exp
[\frac{i}{\hbar} \int L dx
+ \frac{i}{\hbar}J_{\mu}(\bar{\psi} \gamma^{\mu} \psi)] = 0  \ee
and
\be \Ds \int DA D\bar{\psi} D\psi  [D^{-1}_{\mu \nu} A^{\mu} +
e(\bar{\psi} \gamma_{\mu} \psi)]
\exp [\frac{i}{\hbar} \int L dx
+ \frac{i}{\hbar}J_{\mu}(\bar{\psi} \gamma^{\mu} \psi)] = 0  \ee
Thus we have
\be \Ds \langle A_{\mu} \rangle _{J} = e \int dy D_{\mu \nu} (x,y)
\Phi^{\nu}(y), \ee
where $\Phi$ has been defined in (21).

Keeping in mind that
\be J_{\mu} = - \frac{\delta \Gamma}{\delta \Phi ^{\mu}} \ee
we can have from (17), (19), (32) the following equation for the
effective action of the
local composite fields:
\be \ba \Ds \Gamma - \int dx \frac{\delta \Gamma}{\delta
\Phi ^{\mu}} \Phi^{\mu} =
\frac{e^{2}}{2} \int dz_{1} dz_{2} \Phi^{\rho}(z_{1}) D_{\lambda \rho}
(z_{1}, z_{2})\Phi^{\lambda}(z_{2}) -\\
\Ds - i\hbar \Tr \ln T + \hbar^{2} Tr [D^{-1\;\mu \nu} \frac{\delta^{2}}
{\delta \Phi^{\mu} \delta \Phi^{\nu}}
 ( \Tr \ln T )] \ea \ee
where
\be \Ds T = ( i\gamma^{\mu} \partial_{\mu} - m + e^{2} \gamma^{\mu}
\int dy D_{\mu \nu} (x, y) \Phi^{\nu}(y) -
\frac{\delta \Gamma}{\delta \Phi^{\mu}}\gamma^{\mu} ) \delta(x, y) \ee

Thus, starting from the equation (17) for the effective action of
ordinary field we have received
the equation (35) for the effective action of the local composite field.

4. We will seek the solution of (35) in the form
\be \Gamma = \Gamma_{0} + \hbar \Gamma_{1} + \hbar^{2} \Gamma_{2} + \ldots \ee
where $\Gamma_{0}$ is the tree effective action and
$\Gamma_{1}$, $\Gamma_{2}$ are
the quantum corrections.

>From (35) we have for $\Gamma_{0}$
\be \Gamma_{0}[\Phi] = - \frac{e^{2}}{2} \int dx dy \Phi^{\mu}(x) D_{\mu \nu}
(x, y) \Phi^{\nu}(y). \ee

We should point out on the absence of term like $C\Phi$ ($C$ is a constant)
in (37). This is because
we must comply with the condition
\be \Ds \left. \frac{\delta \Gamma}{\delta \Phi} \right |_{J=0} = 0 \ee
which follows from (33)(also $\Phi^{\mu} |_{J=0} = \langle \bar{\psi}
\gamma^{\mu} \psi \rangle = 0$ [11]).
The "classic" field can be derived after substitution (38) into (34)
\be J_{\mu}(x) = e^{2} \int dy D_{\mu \nu}(x, y) \Phi^{\nu}(y) \ee
Thus
\be \Phi^{\mu}(y) = \frac{1}{e^{2}} \int dx D^{\mu \nu \;-1}(x, y)
J_{\nu}(x). \ee

>From (21) we can find the tree expression for $W$
\be W[J] = \frac{1}{2} \int dx dy J_{\mu}(x) [e^{2} D^{\mu \nu}]^{-1}
(x, y) J_{\nu}(y) \ee
and the propagator for the "free" composite field (Fig.1)
\be \Ds \Delta^{\mu \nu} (x,y) = \frac{\delta ^{2} W}{\delta J_{\mu}(x)
\delta J_{\nu}(y)} =
[e^{2} D_{\mu \nu}]^{-1}(x, y) .\ee

Let us turn to calculation of $\Gamma_{1}$. The corresponding equation is
\be \ba \Ds \Gamma_{1}[\Phi] - \int dx \frac{\delta \Gamma_{1}}{\delta
\Phi^{\lambda}} \Phi^{\lambda} =
- i \Tr \ln [1 + 2e^{2} K(x,y) \int dz \gamma^{\mu} D_{\mu \nu} (x,z)
\Phi^{\nu}(z)] - \\
\Ds - i \Tr \ln [K^{-1}(x,y)] \ea \ee
where
\be K(x, y) = [ i\gamma^{\mu} \partial_{\mu} - m ]^{-1} \delta(x, y). \ee
If the both sides of (44) are presented as series over $\Phi$ it can be
easily seen that
the linear term on $\Phi$ is vanishing in the lhs. But in the rhs we will have
\be \int dx 2 e^{2} K (x,x) \int dz \gamma^{\mu} D_{\mu \nu} (x,z)
\Phi^{\lambda}(z) \ee
This is just a QED tadpole because $K$ and $D$ which were defined in (5)
and (45) respectively
are the simple propagators of QED (not "propagators in external field").
So the (46) is actually
zero [11]. If we used any other renormalization condition for which tadpoles
are not equal
to zero then in the renormalized Lagrangian (1) we would have terms of
$A_{\mu}$ in the first
power and consequently there would be non-vanishing terms of $\Phi$
in the first power
in (35) and (44).

After formal integration in (44) we will have the result for $\Gamma_{1}$
\be \ba \Ds \Gamma_{1} = -i \Tr \ln F - \\
\Ds - i \int dx dy 2e^{2} K(x,y) \int dz \gamma^{\alpha}
D_{\alpha \beta}(x,z) \Phi^{\beta}(z) (\ln F)_{y,x} \ea \ee
where
\be F = 1 + 2e^{2} K(x,y) \int dz \gamma^{\mu} D_{\mu \nu} (x,z)
\Phi^{\nu}(z). \ee

Now we can formulate the graph rules for our composite field. The equation
of (47) defines the (non-local) vertexes that describe the interacions of
composite field.
As usual, we define the $n$-point vertex  $G_{i}$ as following
\be \Gamma_{1} = \sum_{i} G_{i} \Phi^{i} \ee
\be G_{i} = \left. \frac{1}{n!} \frac{\delta \Gamma_{1}}{\delta \Phi^{i}}
\right |_{\Phi = 0} \ee

An example (3-point vertex) is given in Fig.2. The propagator of the
composite field (43) and the
vertexes defined in (47) - (50) are the elements of graph technic.
The expansion of $\Gamma$ over
$\hbar$ is no more an expansion on loops maden of propagators of
composite field because our
vertexes derived  from (47) already have $\hbar$ in the first power.
For instance, the graph in
Fig.3 which has one loop, maden of the propagator of composite field,
has the order of $\hbar^{3}$.
It can be easily seen after transformations given in Fig.3. Further,
the transformations given
in Fig.4 show that we can always limit ourselves with the vertexes
of the first order like that
given in Pic.2, because all the vertexes of higher orders can be re-written
in terms of vertexes
given in (50) and the propagator (43). Thus, the order of a graph
on $\hbar$ is equal the number
of loops of propagator (43) plus the number of vertexes $G_{i}$.

We have formulated all the graph rules in terms of propagator
and vertexes of the local
composite field and excluded all the elements of ordinary fields.
It will allow us to
calculate the 1PI (in sense of the local composite field) part of the
effective action of the
order of 2. The matter is, that as was shown in [1] equations for the
effective action can
contain also one-particle reducible contributions (which is not the
case for ordinary fields).
If our graph contained elements of both composite and ordinary fields
it would be difficult
to analyse the one-particle irreducibility.

We will calculate $\Gamma_{2}$ through the direct summing the 1PI
(in sense of the local
composite field) graphs. Accordingly to our rule of calculating of
order of $\hbar$ only
graphs shown in Fig.5 can have the second order on $\hbar$.
Graphs (a) are reducible
on propagator (43). Thus we have
\be \Ds \Gamma_{2} = \sum^{\infty}_{N=2} \Delta (C^{(N)}G_{N}\Phi^{N-2}) \ee
$C^{(N)}$ is a symmetry factor that defines the number of possible
ways to shorten the
pairs of ends in an $N$-point non-local vertex:
\be C^{(N)} = \frac{N!}{2!(N-2)!} = \frac{1}{2} N(N-1). \ee

Remembering (50) we have for $\Gamma_{2}$
\be \Gamma_{2} = \frac{1}{2} \Delta_{xy} \frac{\delta^{2}}{\delta
\Phi_{x} \delta \Phi_{y}}
\sum^{\infty}_{N=0} G_{N} \Phi^{N} = \frac{1}{2} \Delta_{xy} \frac{\delta^{2}}
{\delta \Phi_{x} \delta \Phi_{y}} \Gamma_{1}. \ee

This result is the 1PI (in the sense of propagator of local composite
field) part of
the effective action of the EM current of QED. It can be easily seen
after substitution
(53) in (35) that (33) does not represent solution of the corresponding
part of (35),
i.e. the Legendre transformation (19) does not lead to 1PI (in the sense
of composite field)
generating functional. Therefore we have proved by direct calculations
the statement
of the work [1] (see above).

In order to return to the common graph technic we should transform
the graphs in Fig.5
the same way as it was done in Fig.3.


\bb
\bibitem{lav} Lavrov P 1990 {\sl Teor.math.phys.}
{\bf 82} p 402.
\bibitem{jack} Jackiw R 1974 {\sl Phys.Rev} {\bf D9}
p 1686.
\bibitem{cj} Cornwall J M, Jackiw R and Tomboulis E 1974 {\sl Phys.Rev.}
{\bf D10} p 2428.
\bibitem{km} Kushnir V A, Miransky B A 1989 {\sl Yad. Fizika}
{\bf 50} p 542
\bibitem{mir} Miransky V A 1993 {\sl Int. J.Mod.Phys.} {\bf A8} p135
\bibitem{yok}  Yokojima S 1995 {\sl Phys.Rev.} {\bf D51} p 2996.
\bibitem{iim} Iliopoulos J, Itzikson C and Martin A 1975 {\sl Rev.Mod.Phys.}
{\bf 47} p 165.
\bibitem{dw} DeWitt B S 1965 {\sl Dynamical Theory of Groups and Fields}
(NY: Gordon and Breach).
\bibitem{fm2}  Faizullaev B A and Musakhanov M M 1995 {\sl
Ann.of Phys.} {\bf 241} p 394;
Faizullaev B A, Galkin D V and Garnov S A 1996 {\sl
Talk at 10 Int. Conf. On Problems in QFT, Alushta, 13-18 May} .
\bibitem{ll} Landau and Lifshitz 1989 {\sl Course of Theoretical Physics
vol IV} (Moscow: Nauka).
\bibitem{vas} Vasiliev A N 1980 {\sl Functional methods in QFT and statistics}
(Leningrad: University Press)
\eb


\newpage
\setlength{\unitlength}{3947sp}%
\begingroup\makeatletter\ifx\SetFigFont\undefined%
\gdef\SetFigFont#1#2#3#4#5{%
  \reset@font\fontsize{#1}{#2pt}%
  \fontfamily{#3}\fontseries{#4}\fontshape{#5}%
  \selectfont}%
\fi\endgroup%
\begin{picture}(1425,1614)(1501,-2065)
\thinlines
\put(1200,-950){$\Delta$}
\put(1810,-770){\line(1,0){1000}}
\put(1951,-961){\oval(300,300)[tr]}
\put(1951,-961){\oval(300,300)[tl]}
\put(2251,-961){\oval(300,300)[bl]}
\put(2251,-961){\oval(300,300)[br]}
\put(2551,-961){\oval(300,300)[tr]}
\put(2551,-961){\oval(300,300)[tl]}
\put(2101,-586){\makebox(0,0)[lb]{\smash{\SetFigFont{12}{14.4}
{\rmdefault}{\mddefault}{\updefault}1}}}
\put(1801,-2011){\makebox(0,0)[lb]{\smash{\SetFigFont{12}{14.4}
{\rmdefault}{\mddefault}{\updefault}Fig.1}}}
\put(1501,-961){\makebox(0,0)[lb]{\smash{\SetFigFont{12}{14.4}
{\rmdefault}{\mddefault}{\updefault}=}}}
\put(2926,-961){\makebox(0,0)[lb]{\smash{\SetFigFont{12}{14.4}
{\rmdefault}{\mddefault}{\updefault}=}}}
\put(3200,-900){\line(1,0){1000}}
\put(3200,-1020){\line(1,0){1000}}
\end{picture}

\newpage
\newcounter{cms}
\setlength{\unitlength}{1mm}

\setlength{\unitlength}{3947sp}%
\begingroup\makeatletter\ifx\SetFigFont\undefined%
\gdef\SetFigFont#1#2#3#4#5{%
  \reset@font\fontsize{#1}{#2pt}%
  \fontfamily{#3}\fontseries{#4}\fontshape{#5}%
  \selectfont}%
\fi\endgroup%
\begin{picture}(3833,2556)(976,-2515)
\thinlines
\put(3601,-1711){\oval(300,300)[tl]}
\put(3601,-1711){\oval(300,300)[bl]}
\put(3601,-2011){\oval(300,300)[br]}
\put(3601,-2011){\oval(300,300)[tr]}
\put(3601,-2311){\oval(300,300)[tl]}
\put(3601,-2311){\oval(300,300)[bl]}
\put(2989,-586){\oval(324,300)[tr]}
\put(2989,-511){\oval(276,150)[tl]}
\put(2701,-361){\oval( 40,150)[tl]}
\put(2832,-361){\oval(302,302)[bl]}
\put(2832,-511){\oval( 38,  2)[br]}
\put(2701,-136){\oval(186,300)[br]}
\put(2626,-136){\oval(336,336)[tr]}
\put(2626, 14){\oval(150, 36)[tl]}
\put(4139,-436){\oval(274,150)[tr]}
\put(4139,-511){\oval(326,300)[tl]}
\put(4389,-436){\oval(226,120)[bl]}
\put(4389,-361){\oval(270,270)[br]}
\put(4501,-361){\oval( 46,150)[tr]}
\put(4764,-61){\oval( 74,  6)[tr]}
\put(4764,-286){\oval(526,456)[tl]}
\put(1351,-961){\circle{474}}
\put(3601,-961){\circle{1400}}
\put(1276,-1036){\makebox(0,0)[lb]{\smash{\SetFigFont{12}{14.4}
{\rmdefault}{\mddefault}{\updefault}3}}}
\put(2101,-1036){\makebox(0,0)[lb]{\smash{\SetFigFont{12}{14.4}
{\rmdefault}{\mddefault}{\updefault}=}}}
\put(976,-2461){\makebox(0,0)[lb]{\smash{\SetFigFont{12}{14.4}
{\rmdefault}{\mddefault}{\updefault}Fig.2}}}
\end{picture}



\newpage
\setlength{\unitlength}{3947sp}%
\begingroup\makeatletter\ifx\SetFigFont\undefined%
\gdef\SetFigFont#1#2#3#4#5{%
  \reset@font\fontsize{#1}{#2pt}%
  \fontfamily{#3}\fontseries{#4}\fontshape{#5}%
  \selectfont}%
\fi\endgroup%
\begin{picture}(11642,3114)(968,-3565)
\thinlines
\put(1801,-1336){\oval(1200,750)[tr]}
\put(1801,-1336){\oval(1200,750)[tl]}
\put(1801,-1336){\oval(1050,600)[tr]}
\put(1801,-1336){\oval(1050,600)[tl]}
\put(1801,-1786){\oval(1200,750)[bl]}
\put(1801,-1786){\oval(1200,750)[br]}
\put(1801,-1786){\oval(1050,600)[bl]}
\put(1801,-1786){\oval(1050,600)[br]}
\put(4351,-961){\oval(300,300)[tr]}
\put(4351,-961){\oval(300,300)[tl]}
\put(4651,-961){\oval(300,300)[bl]}
\put(4651,-961){\oval(300,300)[br]}
\put(4951,-961){\oval(300,300)[tr]}
\put(4951,-961){\oval(300,300)[tl]}
\put(4351,-2161){\oval(300,300)[bl]}
\put(4351,-2161){\oval(300,300)[br]}
\put(4651,-2161){\oval(300,150)[tr]}
\put(4651,-2161){\oval(300,150)[tl]}
\put(4951,-2161){\oval(300,300)[bl]}
\put(4951,-2161){\oval(300,300)[br]}
\put(3451,-1336){\oval(300,300)[tr]}
\put(3451,-1336){\oval(300,300)[tl]}
\put(3151,-1336){\oval(300,300)[bl]}
\put(3151,-1336){\oval(300,300)[br]}
\put(3451,-1786){\oval(300,300)[bl]}
\put(3451,-1786){\oval(300,300)[br]}
\put(3151,-1786){\oval(300,300)[tr]}
\put(3151,-1786){\oval(300,300)[tl]}
\put(7051,-961){\oval(300,300)[tr]}
\put(7051,-961){\oval(300,300)[tl]}
\put(6751,-961){\oval(300,300)[bl]}
\put(6751,-961){\oval(300,300)[br]}
\put(6414,-961){\oval(374,308)[tr]}
\put(6414,-961){\oval(376,308)[tl]}
\put(7051,-2161){\oval(300,300)[bl]}
\put(7051,-2161){\oval(300,300)[br]}
\put(6751,-2161){\oval(300,300)[tr]}
\put(6751,-2161){\oval(300,300)[tl]}
\put(6451,-2161){\oval(300,300)[bl]}
\put(6451,-2161){\oval(300,300)[br]}
\put(7951,-1561){\oval(300,300)[tr]}
\put(7951,-1561){\oval(300,300)[tl]}
\put(8251,-1561){\oval(300,300)[bl]}
\put(8251,-1561){\oval(300,300)[br]}
\put(5401,-961){\oval(300,300)[bl]}
\put(5401,-961){\oval(300,300)[br]}
\put(5701,-961){\oval(300,300)[tr]}
\put(5701,-961){\oval(300,300)[tl]}
\put(6001,-961){\oval(300,300)[bl]}
\put(6001,-961){\oval(300,300)[br]}
\put(5326,-2161){\oval(300,300)[tr]}
\put(5326,-2161){\oval(300,300)[tl]}
\put(5626,-2161){\oval(300,300)[bl]}
\put(5626,-2161){\oval(300,300)[br]}
\put(5926,-2161){\oval(300,300)[tr]}
\put(5926,-2161){\oval(300,300)[tl]}
\put(9751,-961){\oval(300,300)[tr]}
\put(9751,-961){\oval(300,300)[tl]}
\put(10051,-961){\oval(300,300)[bl]}
\put(10051,-961){\oval(300,300)[br]}
\put(10351,-961){\oval(300,300)[tr]}
\put(10351,-961){\oval(300,300)[tl]}
\put(10651,-961){\oval(300,300)[bl]}
\put(10651,-961){\oval(300,300)[br]}
\put(11176,-999){\oval(  8, 74)[br]}
\put(10989,-999){\oval(382,384)[tr]}
\put(10989,-961){\oval(376,308)[tl]}
\put(9751,-2161){\oval(300,300)[bl]}
\put(9751,-2161){\oval(300,300)[br]}
\put(10276,-2199){\oval(  8, 74)[br]}
\put(10089,-2199){\oval(382,384)[tr]}
\put(10089,-2161){\oval(376,308)[tl]}
\put(10276,-2199){\oval(  8, 76)[tl]}
\put(10464,-2199){\oval(384,382)[bl]}
\put(10464,-2199){\oval(382,382)[br]}
\put(10651,-2199){\oval(  8, 76)[tr]}
\put(10801,-2161){\oval(300,300)[tr]}
\put(10801,-2161){\oval(300,300)[tl]}
\put(10951,-2180){\oval(  2, 38)[tl]}
\put(11101,-2180){\oval(302,302)[bl]}
\put(11101,-2180){\oval(302,302)[br]}
\put(11251,-2180){\oval(  2, 38)[tr]}
\put(12151,-1561){\oval(300,300)[tr]}
\put(12151,-1561){\oval(300,300)[tl]}
\put(12451,-1561){\oval(300,300)[bl]}
\put(12451,-1561){\oval(300,300)[br]}
\put(8994,-1111){\oval(314,300)[tr]}
\put(8994,-1119){\oval(316,316)[tl]}
\put(8851,-1119){\oval( 30,134)[bl]}
\put(8701,-1186){\oval(300,300)[bl]}
\put(8701,-1186){\oval(300,300)[br]}
\put(8926,-1936){\oval(300,300)[bl]}
\put(8926,-1936){\oval(300,300)[br]}
\put(8639,-1936){\oval(274,150)[tr]}
\put(8639,-2011){\oval(326,300)[tl]}
\put(1201,-1561){\circle{450}}
\put(2401,-1561){\circle{450}}
\put(4201,-1561){\circle{1210}}
\put(5200,-700){\line(1,0){1000}}
\put(5200,-2000){\line(1,0){1000}}
\put(7201,-1561){\circle{1200}}
\put(9601,-1561){\circle{1236}}
\put(11401,-1561){\circle{1210}}
\put(1126,-1636){\makebox(0,0)[lb]{\smash{\SetFigFont{12}{14.4}
{\rmdefault}{\mddefault}{\updefault}4}}}
\put(2326,-1636){\makebox(0,0)[lb]{\smash{\SetFigFont{12}{14.4}
{\rmdefault}{\mddefault}{\updefault}3}}}
\put(5176,-3511){\makebox(0,0)[lb]{\smash{\SetFigFont{12}{14.4}
{\rmdefault}{\mddefault}{\updefault}Fig.3}}}
\put(2701,-1636){\makebox(0,0)[lb]{\smash{\SetFigFont{12}{14.4}
{\rmdefault}{\mddefault}{\updefault}=}}}
\put(8476,-1636){\makebox(0,0)[lb]{\smash{\SetFigFont{12}{14.4}
{\rmdefault}{\mddefault}{\updefault}=}}}
\put(5551,-586){\makebox(0,0)[lb]{\smash{\SetFigFont{12}{14.4}
{\rmdefault}{\mddefault}{\updefault}1}}}
\put(5551,-1786){\makebox(0,0)[lb]{\smash{\SetFigFont{12}{14.4}
{\rmdefault}{\mddefault}{\updefault}1}}}
\end{picture}

\newpage
\setlength{\unitlength}{3947sp}%
\begingroup\makeatletter\ifx\SetFigFont\undefined%
\gdef\SetFigFont#1#2#3#4#5{%
  \reset@font\fontsize{#1}{#2pt}%
  \fontfamily{#3}\fontseries{#4}\fontshape{#5}%
  \selectfont}%
\fi\endgroup%
\begin{picture}(7667,3370)(142,-3790)
\thinlines
\put(7180,-1786){\oval(1158,792)[bl]}
\put(7180,-1561){\oval(1242,1242)[br]}
\put(7180,-1561){\oval(1242,1242)[tr]}
\put(7180,-1336){\oval(1158,792)[tl]}
\put(7177,-1786){\oval(1002,648)[bl]}
\put(7177,-1561){\oval(1098,1098)[br]}
\put(7177,-1561){\oval(1098,1098)[tr]}
\put(7177,-1336){\oval(1002,648)[tl]}
\put(976,-1899){\oval( 12, 74)[br]}
\put(864,-1899){\oval(236,238)[tr]}
\put(864,-1899){\oval(238,238)[tl]}
\put(751,-1899){\oval( 12, 74)[bl]}
\put(1164,-1936){\oval(376,308)[bl]}
\put(1164,-1936){\oval(374,308)[br]}
\put(1651,-1944){\oval( 30,134)[br]}
\put(1509,-1944){\oval(314,316)[tr]}
\put(1509,-1936){\oval(316,300)[tl]}
\put(1201,-811){\oval(300,300)[tl]}
\put(1201,-811){\oval(300,300)[bl]}
\put(1239,-661){\oval( 76, 12)[bl]}
\put(1239,-549){\oval(236,236)[br]}
\put(1239,-549){\oval(236,238)[tr]}
\put(1239,-436){\oval( 76, 12)[tl]}
\put(1801,-1186){\oval(300,300)[tr]}
\put(1801,-1186){\oval(300,300)[tl]}
\put(2101,-1186){\oval(300,300)[bl]}
\put(2101,-1186){\oval(300,300)[br]}
\put(594,-1186){\oval(314,300)[tr]}
\put(594,-1194){\oval(316,316)[tl]}
\put(451,-1194){\oval( 30,134)[bl]}
\put(301,-1261){\oval(300,300)[bl]}
\put(301,-1261){\oval(300,300)[br]}
\put(4201,-811){\oval(300,300)[tl]}
\put(4201,-811){\oval(300,300)[bl]}
\put(4239,-661){\oval( 76, 12)[bl]}
\put(4239,-549){\oval(236,236)[br]}
\put(4239,-549){\oval(236,238)[tr]}
\put(4239,-436){\oval( 76, 12)[tl]}
\put(3601,-1186){\oval(300,300)[tr]}
\put(3601,-1186){\oval(300,300)[tl]}
\put(3301,-1186){\oval(300,300)[bl]}
\put(3301,-1186){\oval(300,300)[br]}
\put(4839,-1111){\oval(374,308)[tr]}
\put(4839,-1149){\oval(384,384)[tl]}
\put(4651,-1149){\oval(  8, 74)[bl]}
\put(5026,-1149){\oval(  8, 76)[tl]}
\put(5214,-1149){\oval(384,382)[bl]}
\put(5214,-1149){\oval(382,382)[br]}
\put(5401,-1149){\oval(  8, 76)[tr]}
\put(3732,-2011){\oval( 38,  2)[tr]}
\put(3732,-2161){\oval(302,302)[tl]}
\put(3732,-2161){\oval(302,302)[bl]}
\put(3732,-2311){\oval( 38,  2)[br]}
\put(3751,-2461){\oval(300,300)[br]}
\put(3751,-2461){\oval(300,300)[tr]}
\put(4651,-2161){\oval(300,300)[br]}
\put(4651,-2161){\oval(300,300)[tr]}
\put(4651,-2461){\oval(300,300)[tl]}
\put(4651,-2461){\oval(300,300)[bl]}
\put(4051,-2836){\oval(300,300)[tr]}
\put(4051,-2836){\oval(300,300)[tl]}
\put(4351,-2836){\oval(300,300)[bl]}
\put(4351,-2836){\oval(300,300)[br]}
\put(1201,-1561){\circle{1210}}
\put(4201,-1561){\circle{1210}}
\put(3700,-2600){\line(1,0){1000}}
\put(6601,-1561){\circle{474}}
\put(1726,-2086){\makebox(0,0)[lb]{\smash{\SetFigFont{12}{14.4}
{\rmdefault}{\mddefault}{\updefault}1}}}
\put(601,-2086){\makebox(0,0)[lb]{\smash{\SetFigFont{12}{14.4}
{\rmdefault}{\mddefault}{\updefault}2}}}
\put(826,-1336){\makebox(0,0)[lb]{\smash{\SetFigFont{12}{14.4}
{\rmdefault}{\mddefault}{\updefault}3}}}
\put(1126,-1111){\makebox(0,0)[lb]{\smash{\SetFigFont{12}{14.4}
{\rmdefault}{\mddefault}{\updefault}4}}}
\put(1501,-1261){\makebox(0,0)[lb]{\smash{\SetFigFont{12}{14.4}
{\rmdefault}{\mddefault}{\updefault}5}}}
\put(2701,-1561){\makebox(0,0)[lb]{\smash{\SetFigFont{12}{14.4}
{\rmdefault}{\mddefault}{\updefault}=}}}
\put(4126,-2536){\makebox(0,0)[lb]{\smash{\SetFigFont{12}{14.4}
{\rmdefault}{\mddefault}{\updefault}1}}}
\put(4576,-1936){\makebox(0,0)[lb]{\smash{\SetFigFont{12}{14.4}
{\rmdefault}{\mddefault}{\updefault}1}}}
\put(3751,-1936){\makebox(0,0)[lb]{\smash{\SetFigFont{12}{14.4}
{\rmdefault}{\mddefault}{\updefault}2}}}
\put(3826,-1336){\makebox(0,0)[lb]{\smash{\SetFigFont{12}{14.4}
{\rmdefault}{\mddefault}{\updefault}3}}}
\put(4201,-1186){\makebox(0,0)[lb]{\smash{\SetFigFont{12}{14.4}
{\rmdefault}{\mddefault}{\updefault}4}}}
\put(4501,-1336){\makebox(0,0)[lb]{\smash{\SetFigFont{12}{14.4}
{\rmdefault}{\mddefault}{\updefault}5}}}
\put(6526,-1636){\makebox(0,0)[lb]{\smash{\SetFigFont{12}{14.4}
{\rmdefault}{\mddefault}{\updefault}5}}}
\put(5701,-1561){\makebox(0,0)[lb]{\smash{\SetFigFont{12}{14.4}
{\rmdefault}{\mddefault}{\updefault}=}}}
\put(3901,-3736){\makebox(0,0)[lb]{\smash{\SetFigFont{12}{14.4}
{\rmdefault}{\mddefault}{\updefault}Fig.4}}}
\end{picture}

\newpage
\setlength{\unitlength}{3947sp}%
\begingroup\makeatletter\ifx\SetFigFont\undefined%
\gdef\SetFigFont#1#2#3#4#5{%
  \reset@font\fontsize{#1}{#2pt}%
  \fontfamily{#3}\fontseries{#4}\fontshape{#5}%
  \selectfont}%
\fi\endgroup%
\begin{picture}(6017,2051)(1493,-2965)
\thinlines
\put(6864,-1861){\oval(1126,676)[bl]}
\put(6864,-1561){\oval(1274,1276)[br]}
\put(6864,-1561){\oval(1274,1274)[tr]}
\put(6864,-1261){\oval(1126,674)[tl]}
\put(6858,-1861){\oval(964,536)[bl]}
\put(6858,-1561){\oval(1136,1136)[br]}
\put(6858,-1561){\oval(1136,1136)[tr]}
\put(6858,-1261){\oval(964,536)[tl]}
\put(1801,-1561){\circle{600}}
\put(2100,-1520){\line(1,0){1200}}
\put(2100,-1600){\line(1,0){1200}}
\put(900,-1700){$\sum^{\infty}_{n,m=1}$}
\put(3601,-1561){\circle{600}}
\put(6301,-1561){\circle{618}}
\put(5400,-1700){$\sum^{n=2}_{\infty}$}
\put(1726,-1636){\makebox(0,0)[lb]{\smash{\SetFigFont{12}{14.4}
{\rmdefault}{\mddefault}{\updefault}n}}}
\put(3526,-1636){\makebox(0,0)[lb]{\smash{\SetFigFont{12}{14.4}
{\rmdefault}{\mddefault}{\updefault}m}}}
\put(6151,-1636){\makebox(0,0)[lb]{\smash{\SetFigFont{12}{14.4}
{\rmdefault}{\mddefault}{\updefault}n+2}}}
\put(2701,-2536){\makebox(0,0)[lb]{\smash{\SetFigFont{12}{14.4}
{\rmdefault}{\mddefault}{\updefault}(a)}}}
\put(7126,-2461){\makebox(0,0)[lb]{\smash{\SetFigFont{12}{14.4}
{\rmdefault}{\mddefault}{\updefault}(b)}}}
\put(4501,-2911){\makebox(0,0)[lb]{\smash{\SetFigFont{12}{14.4}
{\rmdefault}{\mddefault}{\updefault}Fig.5}}}
\end{picture}

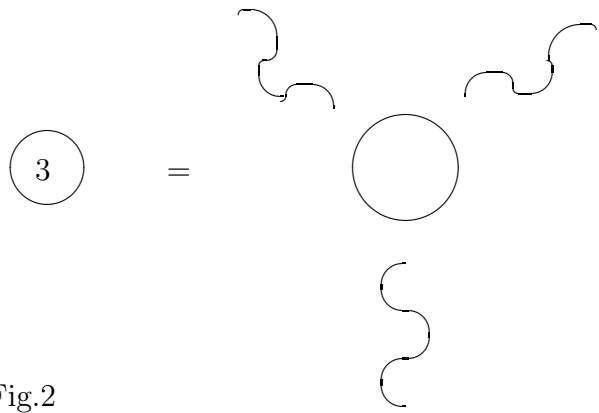
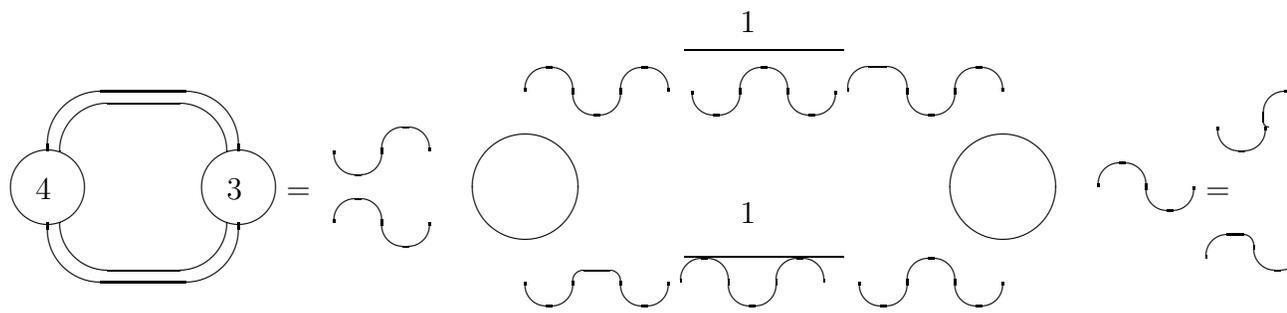
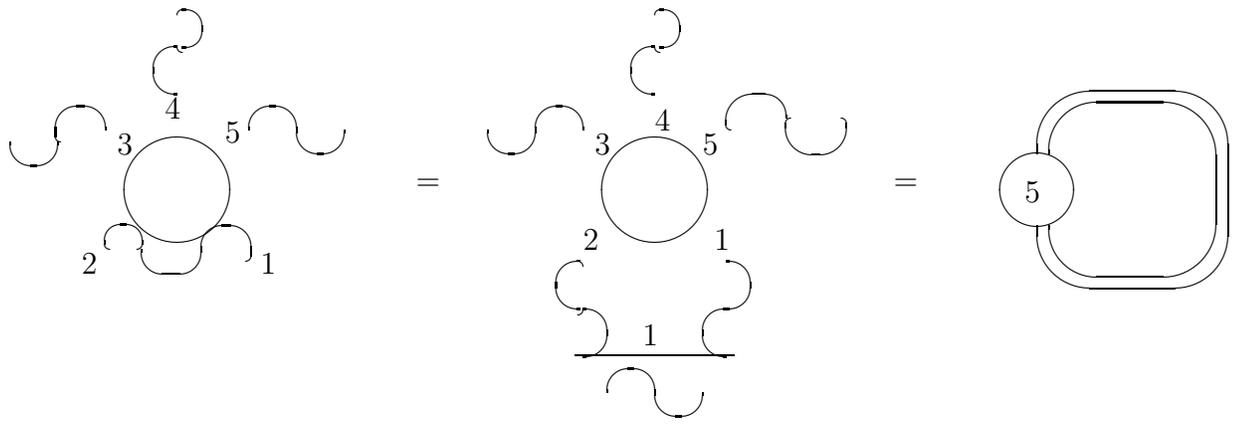
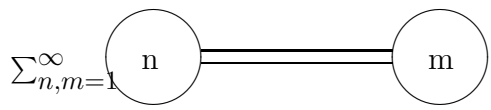
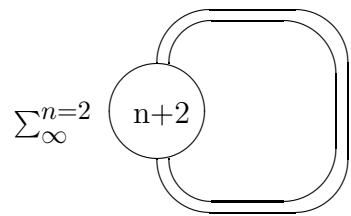
\end{document}